\documentclass[prd,preprint,nofootinbib]{revtex4}
\usepackage{graphicx}

\begin{document}

\newcommand{\gsim}{ \mathop{}_{\textstyle \sim}^{\textstyle >} }
\newcommand{\lsim}{ \mathop{}_{\textstyle \sim}^{\textstyle <} }
\newcommand{\vev}[1]{ \left\langle {#1} \right\rangle }

\newcommand{\bear}{\begin{array}}  \newcommand{\eear}{\end{array}}
\newcommand{\bea}{\begin{eqnarray}}  \newcommand{\eea}{\end{eqnarray}}
\newcommand{\beq}{\begin{equation}}  \newcommand{\eeq}{\end{equation}}
\newcommand{\bef}{\begin{figure}}  \newcommand{\eef}{\end{figure}}
\newcommand{\bec}{\begin{center}}  \newcommand{\eec}{\end{center}}
\newcommand{\non}{\nonumber}  \newcommand{\eqn}[1]{\beq {#1}\eeq}
\newcommand{\la}{\left\langle} \newcommand{\ra}{\right\rangle}
\def\lrf#1#2{ \left(\frac{#1}{#2}\right)}
\def\lrfp#1#2#3{ \left(\frac{#1}{#2}\right)^{#3}}

%

\renewcommand{\thefootnote}{\alph{footnote}}

\renewcommand{\thefootnote}{\fnsymbol{footnote}}
\preprint{DESY 07-156}
\title{Dark Matter in Gauge Mediated Supersymmetry Breaking \\ using Metastable Vacua}
\renewcommand{\thefootnote}{\alph{footnote}}

\author{Motoi Endo and  Fuminobu Takahashi}

\affiliation{
Deutsches Elektronen Synchrotron DESY, Notkestrasse 85,
  22603 Hamburg, Germany }

\begin{abstract}
\noindent
We point out that, in a class of gauge mediation models using
metastable supersymmetry breaking vacua, the minimum of the
supersymmetry breaking field in the early universe is dynamically
deviated from the one in the low energy.  The deviation induces
coherent oscillations of the supersymmetry breaking field, which
decays into the gravitinos.  For certain parameters, it can produce a
right amount of the gravitinos to account for the observed dark
matter.
\end{abstract}


\maketitle

\section{Introduction}
\label{sec:1}
Gauge mediation (GM) of supersymmetry (SUSY) breaking~\cite{GMSB} is a
natural solution to the phenomenological problems such as excessive
flavor-changing neutral currents. In spite of the successes, building
a realistic model is a rather non-trivial task. According to the
argument of Nelson and Seiberg, an exact $U(1)_R$ symmetry is required
if the superpotential is generic~\cite{Nelson:1993nf}.  This
observation strongly limits possible models, and a lot of efforts have
been devoted to building a realistic one.

Recently, Murayama and Nomura has proposed drastically simplified
models, focusing on metastable vacua in the SUSY breaking
sector~\cite{Murayama:2006yf}.  Even though the entire superpotential
does not possess the exact $U(1)_R$ symmetry, an accidental one exists
near a local SUSY breaking minimum~\cite{Intriligator:2006dd}.  Such a
scenario is viable as long as the metastable vacua have a sufficiently
long life time.

In a class of the metastable SUSY breaking models, there are the local
SUSY breaking vacua near the origin of the SUSY breaking fields, where
the accidental $U(1)_R$ symmetry exists. On the other hand, the
breaking of $U(1)_R$ symmetry is necessarily involved in the messenger
sector to mediate the SUSY breaking to the gauginos in the
supersymmetric standard model (SSM).  Since the SUSY breaking field (denoted
by $S$) linearly couples to the messenger fields, the breaking of the
$U(1)_R$ symmetry induces a linear term of $S$ in the K\"ahler
potential.
				   
Such a linear term forces the SUSY breaking field $S$ to deviate from
its minimum in the low energy, while the inflaton field dominates the
energy of the universe.  When the Hubble parameter becomes comparable
to the mass of $S$, it starts to oscillate coherently around the SUSY
breaking vacuum, and then decays into the gravitinos. In the GM
models, the gravitino is stable and behaves as dark matter (DM) in the
universe as long as the R-parity is preserved. Thus, DM is generally
produced from the SUSY breaking field in a class of the GM models
using the metastable vacua. In this letter, we study the gravitino
production in this scheme.

Before proceeding to the details, let us comment on the previous
works. The cosmological evolution of the SUSY breaking field has been
studied in the models with the metastable SUSY breaking 
vacua in Refs.~\cite{Abel:2006cr}. Those literatures assumed so high reheating 
temperature as to make the SUSY breaking sector to be in thermal equilibrium.
Then the gravitino is expected to reach thermal equilibrium, and as a result,
its abundance exceeds the observed DM abundance or it erases the
density fluctuation too much, unless the gravitino mass is smaller than
$16$\,eV~\cite{Viel:2005qj}. For a wide range of the gravitino mass, i.e., from $16$\,eV
to $O(10)$\,GeV, therefore, the SUSY breaking sector should not be thermalized
as long as the standard thermal history of the universe is assumed.
In this letter, we do not pursue this possibility and assume that the SUSY
breaking sector never reaches thermal equilibrium. 
 On the other hand, Ibe and Kitano discussed the
gravitino production from the SUSY breaking field in the GM models
using the metastable vacua~\cite{Ibe:2006rc}.  They assumed a different
thermal history taking a different set of the parameters; 
in particular, the reheating temperature in our scenario is as high as
$10^{8-10}\,$GeV, while they considered the low reheating temperature.
Furthermore, the interaction that shifts the SUSY breaking field from its
minimum in the low energy is different from that considered in Ref.~\cite{Ibe:2006rc}.

\section{Model}
\label{sec:2}
In this section we provide a model of the gauge mediation using the
metastable vacua. To be explicit, we adopt a model by Murayama and
Nomura given in Ref.~\cite{Murayama:2006yf}. The K\"ahler potential
and the superpotential for a gauge singlet chiral field $S$ and the
messengers $f$ and $\bar{f}$ are written as
\bea
\label{K}
K &=& |S|^2 - \frac{|S|^4}{4 \Lambda^2} + |f|^2 + |\bar{f}|^2,\\
\label{W}
W &=& -\mu^2 S + \kappa S f \bar{f} + M f \bar{f},
\eea
where the higher order corrections of $O(|S|^6/\Lambda^4)$ are omitted
for simplicity in the K\"ahler potential. In the following we take $f$
and $\bar{f}$ to be in ${\bf 5} + {\bf 5^*}$ representation of
$SU(5)_{\rm GUT}$, and $\mu^2$, $\kappa$ and $M$ are set to be real
and positive without loss of generality. We assign the charges
$U(1)_R$ symmetry as $R[S]=2$ and $R[f]=R[\bar{f}]=0$. Then one can
see that the messenger mass term explicitly violates the $U(1)_R$
symmetry to $Z_2$.

For our purpose we do not need to specify the UV physics above a scale
$\Lambda$ that provides the second term in Eq.~(\ref{K}) as well as
the first term in Eq.~(\ref{W}).  We simply note here that there are
many explicit models that actually lead to this low energy effective
theory. (See Ref.~\cite{Murayama:2006yf} for examples.) In particular,
we do not give the SUSY breaking mechanism explicitly here, which is
assumed to be such that the first term in Eq.~(\ref{W}) is somehow
produced.

From the K\"ahler potential and the superpotential given above, one can show
that there is a SUSY minimum at
\beq
S \;=\; -\frac{M}{\kappa},~~~f \;=\; {\bar f} \;=\; \frac{\mu}{\sqrt{\kappa}}.
\label{susymin}
\eeq
On the other hand,  SUSY is broken at  $S = f = {\bar f} =0$, 
which is a metastable local minimum as long as
\beq
M^2 \;>\; \kappa \mu^2
\label{eq:tachyon}
\eeq
is satisfied, since otherwise one of the messenger scalars becomes
tachyonic.  Note that the second term in Eq.~(\ref{K}) produces a
positive mass of $S$, $m_S = \mu^2/\Lambda$, around the origin.  The
SUSY breaking scale is dictated by the first term in Eq.~(\ref{W}),
and the $F$-term of $S$ is given by $F_S \simeq \mu^2$. Requiring a
vanishing cosmological constant, we can relate 
the SUSY breaking scale to the gravitino mass $m_{3/2}$ as
\beq
\mu \; = \; \left(\sqrt{3}\, m_{3/2} M_P\right)^\frac{1}{2}
 \simeq 2 \times 10^9{\rm\,GeV} \lrfp{m_{3/2}}{1{\rm\,GeV}}{\frac{1}{2}},
\label{mu}
\eeq
where  $M_P = 2.4 \times 10^{18}{\rm \,GeV}$ is the reduced Planck scale.
The SUSY breaking effects are transmitted to the visible sector by the
messenger loops.  The integration of the messengers give rise to the
gaugino masses as~\cite{GMSB}
\begin{equation}
      m_i \simeq \frac{\alpha_i}{4\pi} \frac{\kappa \mu^2}{M} ~~~~~~~{\rm for}~~i=1,2,3.
      \label{eq:m_i}
\end{equation}
Here, $m_{1,2,3}$ and $\alpha_{1,2,3}$ are the gaugino masses and the
gauge coupling constants for $U(1)_Y, SU(2)_L$ and $SU(3)_C$ in the
SSM. We have used the $SU(5)_{\rm GUT}$ normalization for the $U(1)_Y$
gauge coupling constant.  We can express $\kappa \mu^2/M$ in terms of
the gluino mass $m_3$:
\beq
\frac{\kappa \mu^2}{M} \;\simeq\; 1 \times 10^2 {\rm\,TeV} 
\lrfp{\alpha_3}{0.1}{-1} \lrf{m_3}{1{\rm TeV}}.
\label{kaM}
\eeq
From (\ref{mu}) and (\ref{kaM}), one can express $M$ as
\beq
M \;\simeq\; 3 \times 10^{13} {\rm\, GeV}\, \kappa \lrf{\alpha_3}{0.1} 
\lrfp{m_3}{1{\rm TeV}}{-1} \lrf{m_{3/2}}{1{\rm GeV}}.
\label{M}
\eeq
We also assume $m_S \lesssim \Lambda$, or equivalently, 
\beq
\mu \;\lesssim \; \Lambda,
\label{muLambda}
\eeq
since we consider the dynamics of $S$ (e.g. coherent oscillations and
decay), which should be described within the low energy effective theory.  Using
(\ref{muLambda}), we obtain an upper-bound on $m_S$,
\bea
m_S \;=\; \frac{\mu^2}{\Lambda} &\simeq& 
40{\rm\,TeV} \lrf{m_{3/2}}{1{\rm\,GeV}} \lrfp{\Lambda}{10^{14}{\rm GeV}}{-1},\non\\
&\lesssim&
2 \times 10^9{\rm\,GeV} \lrfp{m_{3/2}}{1{\rm\,GeV}}{\frac{1}{2}}.
\label{ms}
\eea

Lastly, let us discuss radiative corrections to the K\"ahler
potential.  Integrating the messenger loop, the relevant corrections
are given by
\bea
K^{(1)} &=& K^{(1)}_{nh} + K^{(1)}_{h},
\eea
with
\bea
\label{nh}
K^{(1)}_{nh} &=& - \frac{5 M^2}{16 \pi^2} \left\{
\frac{1}{2} \lrfp{\kappa}{M}{3} |S|^2 (S+S^\dagger) -\frac{1}{6} \lrfp{\kappa}{M}{4}
|S|^2 (S^2 + S^{\dagger 2}) + \cdots \right\},\\
\label{h}
K^{(1)}_{h} &=&-\frac{5 M^2 }{16 \pi^2}  \left\{ \frac{\kappa}{M} (S + S^\dag)
+ \cdots \right\},
\label{linear}
\eea
where we have separated the holomorphic terms and the non-holomorphic
ones.  Note that $S$ is assumed to be much smaller than $M/\kappa$ so
that $S$ sits far away from the SUSY minimum (see (\ref{susymin})).
As explained in Introduction, the reason why such corrections appear
is that the $R$-symmetry is explicitly broken by the messenger mass
term.

One can check that the radiative corrections (\ref{nh}) reproduce the
result of the Coleman-Weinberg potential for $S$ given in
Ref.~\cite{Murayama:2006yf}, up to the order explicitly shown in
(\ref{nh}):
\beq
V^{(1)}_{nh} \;=\; \frac{5 \mu^4}{16 \pi^2}\left\{ \frac{\kappa^3}{M} (S+S^\dagger)
- \frac{\kappa^4}{2M^2} (S^2+S^{\dagger 2}) + \cdots \right\}
\eeq
To avoid the mass of $S$ to become tachyonic due to the radiative corrections,
we require~\cite{Murayama:2006yf}
\beq
M \;\gtrsim\; \frac{\kappa^2}{4\pi} \Lambda,
\eeq
throughout this letter.

As far as the SUSY breaking sector is concerned, the linear term in
the K\"ahler potential (\ref{h}) does not modify the scalar potential
significantly.  In the very early universe, however, such a linear
term makes the minimum of the scalar potential to deviate from the
origin. Therefore, it is crucial for cosmological evolution of $S$ to
take into account the linear term in the K\"ahler potential, as we
will show in the next section.

\section{Cosmology}
\label{sec:3}
Now we consider the cosmological evolution of the SUSY breaking field,
$S$.  First let us give a sketch how $S$ is deviated from the origin
due to the linear term in the K\"ahler potential, and estimate the
cosmic abundance.  While the $F$-term of the inflaton dominates the
universe, the scalar potential of $S$ is approximately given by~\footnote
{
The linear term of $S$ generically appear in the scalar potential due to the 
supergravity effects~\cite{Ibe:2006rc}. One can neglect
its effect on the dynamics of $S$, as long as
$\kappa \gtrsim 0.04\, (\alpha_3/0.1)^{1/2} (1{\rm TeV}/m_3)^{1/2} (m_{3/2}/1{\rm GeV})^{1/2}
(\Lambda/10^{14}{\rm GeV})$. We assume that this inequality is satisfied in the following
analysis.
}
\bea
V(S) &\simeq& e^{K} (3 H^2 M_P^2),\non\\
	  &\simeq& 3 H^2 \left( |S|^2 - \frac{5 \kappa}{16 \pi^2} M(S+S^\dagger) + \cdots\right),
\label{V}	  
\eea
where we have assumed that $S$ does not couple to the inflaton in the
K\"ahler potential for simplicity~\footnote{
Even in the presence of the interactions, the following argument does not
change qualitatively.
}.  The scalar potential has a minimum
given by
\beq
S_c \;= \; \frac{5 \kappa}{16 \pi^2} M.
\label{S}
\eeq
If this minimum $S_c$ exceeds $\Lambda$, there is no stable minimum in
the low-energy theory in the early universe. Depending on the UV
theory, the system may settle down at the SUSY minimum  in
this case. To avoid such a situation, 
we impose $S_c < \Lambda $ in the following.

When the Hubble parameter becomes comparable to the mass of $S$,
it starts to oscillate around the minimum, $S \simeq 0$, with an initial
amplitude $S_c$. The abundance of $S$ is estimated as
\bea
\frac{n_S}{s} &\simeq& \frac{3 T_R}{4} \left( \frac{5 \kappa}{16 \pi^2} \right)^2
				\frac{m_S M^2}{3 m_S^2 M_P^2},\non\\
&\simeq&1 \times 10^{-10}\, \kappa^4 \lrfp{\alpha_3}{0.1}{2} \lrfp{m_3}{1{\rm TeV}}{-2} 
			\lrf{m_{3/2}}{1{\rm GeV}} 
	\lrf{T_R}{10^8 {\rm GeV}}\lrf{\Lambda}{10^{14} {\rm GeV}},
\label{nss}				
\eea
where $n_S$ is the number density of $S$, $s$ is the entropy density,
$T_R$ denotes the reheating temperature, and we have used (\ref{M}) to
eliminate $M$.  We have assumed here that the reheating has not
completed when $H = m_S$. This assumption is indeed reasonable, since
otherwise too many gravitinos are produced by thermal scatterings in
plasma.

Several comments are in order. First, we have assumed that the Hubble
parameter during inflation, $H_I$, is larger than the mass of $S$,
i.e., $H_I > m_S$.  If $m_S$ is larger than $H_I$, the deviation of
$S$ is suppressed by $(H_I/m_S)^2$.  Considering the upper-bound on
$m_S$ given by (\ref{ms}), however, the assumption $H_I > m_S$ is
valid except for low-scale inflation models. Second, although our model is
given only below the scale $\Lambda$, this does not limit the
application of the above arguments only to the inflation models with
$H_I < \Lambda$.  For $H_I < \Lambda$, the above scalar potential
(\ref{V}) is obviously valid both during and after inflation. On the
other hand, for $H_I > \Lambda$, one cannot use (\ref{V}) during
inflation. After inflation, however, the Hubble parameter decreases
and becomes smaller than $\Lambda$ at certain point. If the system can
be then described by the low energy effective theory, $S$ will quickly
settle down at the potential minimum (\ref{S})~\footnote{
Depending on the details of the UV theory, it is possible that the position of $S$
is larger than $\Lambda$ and the system cannot be described by the model given 
by (\ref{K}) and (\ref{W}) even for $H < \Lambda$. Then, the abundance of $S$ 
will generically become larger than (\ref{nss}), and so, our estimate is conservative.
}.

Next let us consider the decay processes of $S$.  The SUSY breaking
field $S$ will decay into a pair of the gravitinos, and the decay rate
is~\cite{Casalbuoni:1988kv,Endo:2006zj}
\beq
\Gamma_{3/2} \;\simeq \; \frac{1}{96 \pi} \frac{m_S^5}{m_{3/2}^2 M_P^2} 
\simeq \frac{1}{32 \pi} \frac{m_S^3}{\Lambda^2},
\label{g32}
\eeq
where we have used (\ref{mu}). If $m_S$ is smaller than $2 M$, 
$S$ decays into the gauge sector through the messenger
loop~\footnote{
The decay into the scalars does not change our argument significantly.
}. The relevant interactions are
\begin{eqnarray}
    \mathcal{L}\; \simeq\;
    -\frac{\alpha_i}{4\pi} \frac{\kappa}{M} S
    \bigg[ 
    - \frac{1}{4} F_{\mu\nu}^{(i)} F^{(i) \mu\nu} 
    + \frac{i}{8} \epsilon^{\mu\nu\rho\sigma} 
      F_{\mu\nu}^{(i)} F_{\rho\sigma}^{(i)} 
    - \frac{\kappa F_S}{ M} 
      \bar{\lambda}^{(i)} {\mathcal P}_L \lambda^{(i)} 
    \bigg] + {\rm h.c.},
\end{eqnarray}
where we neglected terms with higher orders of $\kappa \langle S \rangle/M$. 
In particular, $S$ decays into the gluons and gluinos, and the decay rates are
\beq
\Gamma_g \;\simeq\; \frac{\alpha_3^2 \kappa^2}{64 \pi^3} \frac{m_S^3}{M^2}
\label{gg}
\eeq
and~\cite{Ibe:2006rc}
\beq
\Gamma_{\tilde g} \;\simeq\; \frac{\kappa^2 }{\pi} \frac{m_3^2 \,m_S}{M^2},
\label{ggluino}
\eeq
respectively. Note that the decay rate into the gluinos is smaller than
that into the gluons, if $m_S$ is much larger than the gluino mass~\cite{Endo:2006ix},
\beq
m_S \;\gg\; \frac{8 \pi}{\alpha_3} m_3,
\eeq
and vice versa.  Using (\ref{g32}), (\ref{gg}) and (\ref{ggluino}),
we obtain the branching ratio of the gravitino production,
\beq
B_{3/2} \;\simeq\; \frac{1}{1 + r}
\eeq
with
\beq
r \;\equiv\; \frac{8}{3}  \lrfp{m_3 \,\Lambda}{m_{3/2} M_P}{2}+ 
\frac{32}{9} \lrfp{4 \pi}{\alpha_3}{2} \lrfp{m_3 \,\Lambda}{m_{3/2} M_P}{4}.
\eeq
Therefore, the SUSY breaking field will dominantly decay into the
gravitinos if $r \lesssim 1$, or equivalently,
%
\beq
\Lambda \;\lesssim\; 
2 \times 10^{14} {\rm\, GeV}\,\lrfp{\alpha_3}{0.1}{\frac{1}{2}} \lrfp{m_3}{1{\rm TeV}}{-1}
\lrf{m_{3/2}}{1{\rm GeV}}.
\eeq
We will assume this inequality is met in the following analysis.  The
decay temperature of $S$ is given by
\bea
T_d &\equiv& \lrfp{\pi^2 g_*}{10}{-\frac{1}{4}} \sqrt{\Gamma_{3/2} M_P},\non\\
	&\simeq&4 {\rm\,GeV}\,\lrfp{g_*}{100}{-\frac{1}{4}} \lrfp{m_{3/2}}{1{\rm GeV}}{\frac{3}{2}} 
			     \lrfp{\Lambda}{10^{14}{\rm GeV}}{-\frac{5}{2}},
\eea
where $g_*$ counts the relativistic degrees of freedom at the decay.

Now we can estimate the gravitino abundance. Since $S$ dominantly
decays into a
pair of the gravitinos, the gravitino abundance is given by
\beq
Y_{3/2} \;\simeq \;2 \times 10^{-10}\, \kappa^4 \lrfp{\alpha_3}{0.1}{2} \lrfp{m_3}{1{\rm TeV}}{-2} 
			\lrf{m_{3/2}}{1{\rm GeV}} \lrf{T_R}{10^8 {\rm GeV}}\lrf{\Lambda}{10^{14} {\rm GeV}}.
\eeq
The density parameter of the gravitino is
\beq
\Omega_{3/2} h^2 \;\simeq\; 0.06\, \kappa^4 \lrfp{\alpha_3}{0.1}{2} \lrfp{m_3}{1{\rm TeV}}{-2} 
			\lrf{m_{3/2}}{1{\rm GeV}} \lrf{T_R}{10^8 {\rm GeV}}\lrf{\Lambda}{10^{14} {\rm GeV}}.
\eeq
where $h$ is the present Hubble parameter in units of 100 km/s/Mpc.
Therefore, a right amount of the gravitinos can be produced by the
decay of the SUSY breaking field, $S$. Note that, for the non-thermally
produced gravitinos to be a dominant component of DM, $\kappa$ 
should not be suppressed.  The reason is as follows. 
For a small $\kappa$, the messenger mass
is also small to keep the size of the soft masses in the SSM sector (see
(\ref{M})). Since the shift of the $S$ field is proportional to the breaking
of $U(1)_R$ symmetry, i.e., the messenger mass, the gravitino abundance
is suppressed for a small value of $\kappa$. Furthermore, depending on the reheating 
temperature and the mass spectrum of the SSM particles, the thermal production 
of the gravitinos and the NLSP decay may also give sizable 
contributions~\cite{Moroi:1993mb,Bolz:1998ek,Bolz:2000fu,Ellis:2003dn,Pradler:2006qh}.   
Note also that the relatively high reheating temperature is favored,
which may accommodate the thermal leptogenesis scenario~\cite{Fukugita:1986hr}.

Finally, let us estimate the free streaming length of the gravitinos
produced by the decay of $S$.  The comoving free streaming length
$\lambda_{FS}$ at matter-radiation equality is defined by
\beq
\lambda_{FS} \;\equiv \; \int_{t_D}^{t_{\rm eq}} \frac{v_{3/2}(t)}{a(t)} dt,
\label{eq:fs}
\eeq
where $a(t)$ is the scale factor, and $t_D$ and $t_{\rm eq} (\sim 2
\times 10^{12}{\rm\, sec})$ denote the time at the $S$ decay and
at matter-radiation equality, respectively.  $v_{3/2}$ is the velocity
of the gravitino, given by
\beq
\displaystyle{v_{3/2}(t) \;=\; \frac{|{\bf p}_{3/2}|}{E_{3/2}}
 \simeq \frac{\frac{m_S}{2} \lrf{a_D}{a(t)}}{\sqrt{m_{3/2}^2 
 + \frac{ m_S^2}{4} \lrfp{a_D}{a(t)}{2}}}},
\eeq
where we have approximated $m_S \gg m_{3/2}$, and 
$a_D$ is the scale factor at the  decay of $S$.  Integrating
(\ref{eq:fs}) yields
\bea
\lambda_{FS} &\simeq& \frac{1}{H_0 \sqrt{1 + z_{\rm eq}}} X^{-1} \sinh^{-1} X,\non\\
&\sim&1 {\rm\, kpc}\,   \lrfp{g_*}{100}{\frac{1}{2}}   \lrfp{m_{3/2}}{1{\rm GeV}}{-\frac{3}{2}}                  				    \lrfp{\Lambda}{10^{14} {\rm GeV}}{\frac{3}{2}}
\label{eq:lambdafs}				    
\eea
with
\bea
X &\equiv& \frac{2 m_{3/2}}{m_S} \frac{a_{\rm eq}}{a_D},\non\\
   &\sim& 10^6\,  \lrfp{g_*}{100}{-\frac{1}{2}}    	   
     \lrfp{m_{3/2}}{1{\rm GeV}}{\frac{3}{2}} 
     \lrfp{\Lambda}{10^{14}{\rm GeV}}{-\frac{3}{2}},
\label{eq:X}				    
\eea
where $H_0$ is the Hubble parameter at present, and $z_{\rm eq}$ and
$a_{\rm eq}$ are the red-shift and the scale factor at the
matter-radiation equality. In the second equation of
(\ref{eq:lambdafs}), we have used $H_0^{-1} \sim 4 \times 10^3 {\rm\,
Mpc}$ and $z_{\rm eq} \sim 3000$. Thus, the free streaming length
$\lambda_{FS}$ is expressed in terms of the gravitino mass and the
scale $\Lambda$, and it can be as small as $1\,$kpc. Interestingly,
the recent observations on the dSph galaxies seem to exhibit a sharp
cut-off around $100\,$pc~\cite{Gilmore:2007fy} in the smallest size of
the galaxies, which may be explained by DM with free
streaming length of $O(100 {\rm \,pc})$~\footnote{It is also possible
to produce a right amount of the gravitino DM from the inflaton
decay~\cite{Takahashi:2007tz}, and the gravitino can have right free
streaming length to explain the cut-off in the smallest size of the
galaxies.}.  Our scenario may be supported by further observations in
the near future.

\section{Conclusions and discussion}
\label{sec:4}
The metastable SUSY breaking vacua provide a drastically
simplified scheme of the gauge mediation. The models possess
an accidental $U(1)_R$ symmetry, which is broken in the
messenger sector. The breaking induces a linear term of the
SUSY breaking field in the K\"ahler potential. We have pointed out that 
the SUSY breaking field is forced away from its minimum due to this
linear term while the  inflaton field dominates the energy of the universe. 
Then the gravitinos are produced when it decays, and we have shown
that a right abundance of the gravitino DM can be realized for certain 
parameters. Further, the free streaming length of the gravitino may explain
the recent observations on the smallest size of the dSph galaxies.

\section*{Acknowledgment}
We thank W. Buchm\"uller for reading the manuscript and comments.


\end{document}